\documentclass[final,5p,times,twocolumn]{elsarticle}

\usepackage{color}
\usepackage{amsmath,amssymb}
\usepackage{graphicx}
\usepackage{subfigure}
\usepackage{dcolumn}
\usepackage{bm}

\newcommand{\Fig}[1]{Figure~\ref{fig:#1}}

\begin{document}

\begin{frontmatter}

\title{Chains of benzenes with lithium-atom adsorption: Vibrations and spontaneous symmetry breaking}

\author[1]{Yenni P. Ortiz}
\author[1]{Thomas Stegmann}
\author[2]{Douglas J. Klein}
\author[1,3]{Thomas H. Seligman}

\address[1]{Instituto de Ciencias F\'isicas, Universidad Nacional Aut\'onoma de M\'exico,
  Avenida Universidad s/n, 62210 Cuernavaca, M\'exico}
\address[2]{Texas $\&$ M University at Galveston, TX 77551, USA}
\address[3]{Centro Internacional de Ciencias, 62210 Cuernavaca, M\'exico}

\begin{abstract}
  We study effects of different configurations of adsorbates on the vibrational modes as well as
  symmetries of polyacenes and poly-\textit{p}-phenylenes focusing on lithium atom adsorption. We
  found that the spectra of the vibrational modes distinguish the different configurations. For more
  regular adsorption schemes the lowest states are bending and torsion modes of the skeleton, which
  are essentially followed by the adsorbate. On poly-\textit{p}-phenylenes we found that lithium
  adsorption reduces and often eliminates the torsion between rings thus increasing symmetry. There
 is spontaneous symmetry breaking in poly-\textit{p}-phenylenes due to double adsorption of
  lithium atoms on alternating rings.
\end{abstract}

\begin{keyword}
conjugated carbon structures, spontaneous symmetry breaking
\end{keyword}

\end{frontmatter}

\section{Introduction}

Adsorption of lithium atoms and other alkali metals to conjugated carbon systems has shown the
possibility of strong deformations which may occur as spontaneous symmetry breaking
\cite{mosh,Ortiztesis,JOS,OStbpublished, JS}. Polyacenes have served as paradigmatic examples
\cite{mosh, Ortiztesis, JS} but similar effects have previously been discussed in other aromatic
molecules \cite{JOS, JS}. The question as to what extent such adsorption can be detected leads to a
comparison between the vibrational spectra of the corresponding molecules. One focus here is on
polyacenes, in particular on anthracene for two reasons: First, it is in this context that
interesting near-periodic structures are seen which are relevant to narrow graphenic nano-ribbons
\cite{OStbpublished}; Second, the matter of spontaneous symmetry breaking is quite clear in these
structures, as we shall see. Further, another type of benzenoid species is studied, namely the
poly-\textit{p}-phenylenes (PPP) we find that adsorption here increases certain (distortive)
similarities to polyacenes, in that torsion present in the naked PPP is suppressed. Notably PPPs are
less chemically reactive (and so much easier to handle) as compared to polyacenes. Also their
potential for branching allows a richer spectrum of future extension of the present work.

The question arises as to how such deformations affect the stiffness of the molecules and to what
extent low-lying modes couple movements of the underlying chains and movements of adsorbates. We
study Li adsorption, manifesting interesting effects: spontaneous symmetry breaking and fairly large
charge transfers. The extension to linear PPPs is very suggestive because they represent the
narrowest "armchair'' graphene boundary \cite{edg1,edg2}. One great difference is that (neutral)
PPPs are not flat and thus seem to be far from armchair edge nano-strips but interestingly we find
that they flatten when subject to adsorption of lithium atoms.

We use DFT calculations to obtain the pertinent data from which we obtain the structure and further
data that determine the vibrational modes. We then proceed to calculate the vibrational modes. Next
we discuss the differences in the spectra and observe that with few exceptions the actual dynamics
of low lying frequencies is much the same as that of the underlying spectrum of the simple (naked)
polyacenes. The purpose is to show how things change with heavier and larger adsorbates.

The need to analyze vibrational modes of such systems naturally arises to enhance structural
understanding. On one hand we would like to know if the spectra are sufficiently different to allow
the identification of a given configuration and on the other hand, whether the structure of low
lying modes indicates some decoupling of the movement of the naked molecule (skeleton) from the
movement of the adsorbates. If indeed such a decoupling happens it would be important to know which
modes are lower. We present this analysis in some detail for the case of anthracene, to find that
the spectra are quite different; furthermore we find the usual scenario, the first few modes are
vibrations of the skeleton which are essentially followed by the adsorbates, though exceptions
occur. These results will lead us to a speculative discussion of the possibility to use such chains,
most likely PPP's, with regularly spaced adsorbates as substrate for quantum register using the
adsorbates as the string of qubits and the vibrations as extended qubits to mediate the information
for control gates in the spirit of the Innsbruck quantum computer \cite{Innsbruck,
  Innsbruck2}. There an ion chain in a casi 1-D trap serves as qubit chain and the collective motion
of the ions to mediate information for control gates. We will not discuss the quantum optical
necessities to implement such a scheme coupling the ions by molecular vibrations and the qubits by
adsorbed Li atoms. Rather we simply point out that a possibly viable structure exists. In particular
we present some configurations of polyacenes and PPP's that fulfill the obvious criteria we need:
Low lying largely decoupled modes of movement of the skeleton.

For PPPs torsion modes will also appear in the lower part of the spectra, and they similarly tend to
decouple at low frequency from the other modes, this have been seen in other carbon systems
\cite{ortizklein5}.  Again there are exceptions. Nevertheless, the prevailing feature of decoupling
of modes for the lowest states is interesting and may be relevant in various aspects not presently
obvious.

We proceed as follows: First we discuss briefly the computational methods used. We then proceed to
discuss anthracene in detail analyzing 14 configurations with zero up to four lithium atoms
adsorbed. Some aspects relating to spontaneous symmetry breaking in infinite systems and their
relation to the Peierls transition are commented as supplement.  Finally we give some conclusions
and an outlook.

\begin{figure*}[htb]
  \centering
  \includegraphics[scale=0.4]{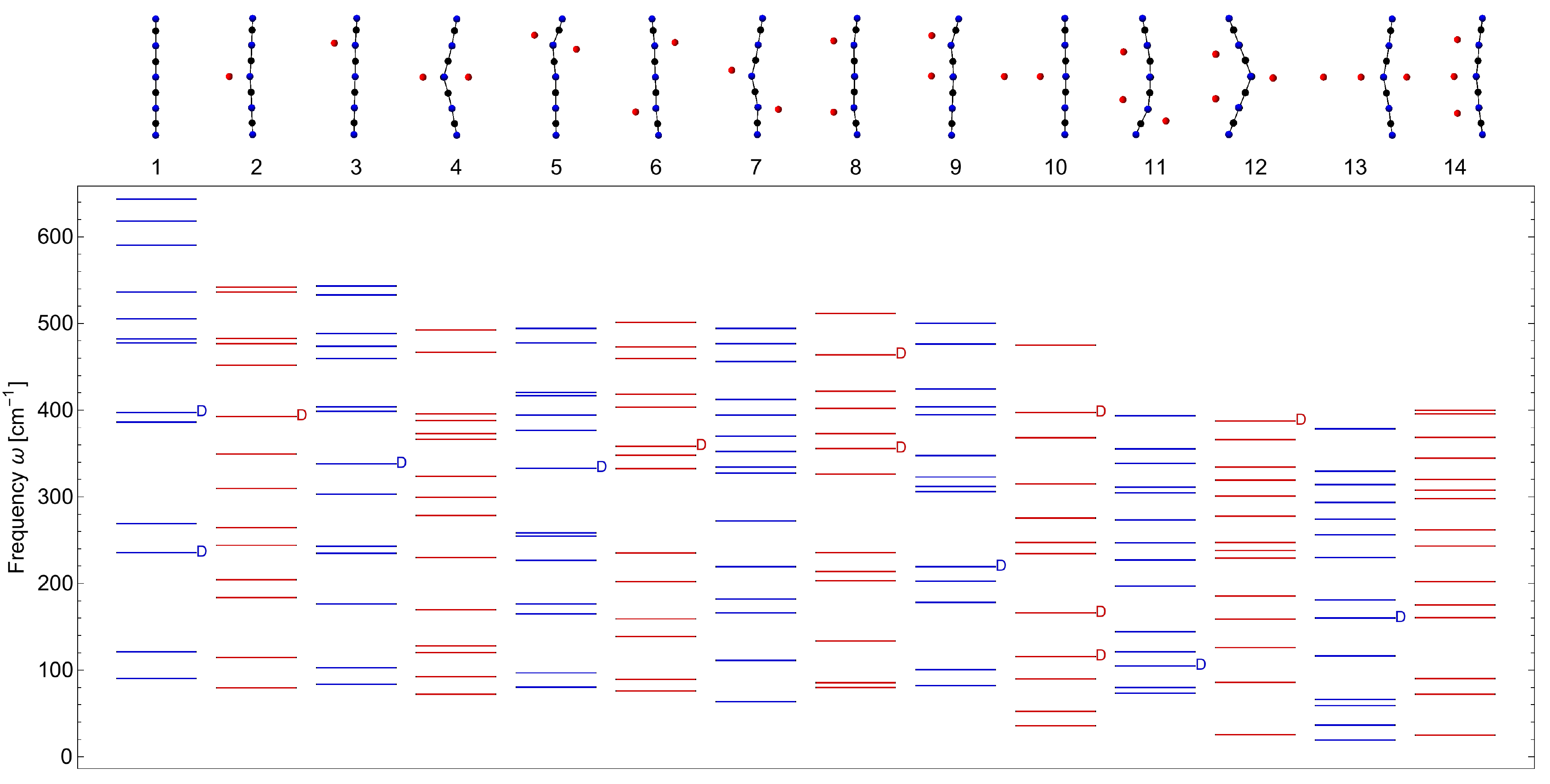}
  \caption{Vibrational spectra of anthracene and anthracene with Li adsorption. In this figure and
    in all following vibrational spectra, 15 modes are shown for each molecule. Tight doublets are
    indicated by the letter D to make sure that they can be identified.}
  \label{fig:1}
\end{figure*}

\section{Computational methods}

The calculations for individual molecules were performed with the GAUSSIAN09 program codes
\cite{gaussian}. Geometry optimizations were made with DFT \cite{dft} using hybrid functional B3LYP
\cite{b3lyp}.  We have selected the 6-311g \cite{basis} basis sets for electronic structure
calculations that use linear combinations of gaussian functions to form the orbitals. Additionally,
we use the improvement of these basis sets by adding d functions to carbon atoms.

To test whether polarization is important we rerun some of the calculations with two different basis
sets that include such terms, 6-31g+(d) and 6-311++g(d,p). Also we used alternatively the
functionals MPW1PW91 \cite{mpw1,mpw2,mpw3} and B3PW91 \cite{b3lyp,mpw1,b3pw91_2}. Those functionals
have been used in carbons systems \cite{C1,C2} as well as in alkaline molecules \cite{alk} which
have been reported to yield very accurate vibrational frequencies and IR intensities.  No
qualitatively difference could be seen in the results, both for the structure and the vibrational
spectra, therefore we do not show the corresponding figures.

The frequency calculations were made with the same package, by determining the second derivatives of
the energy with respect to the Cartesian nuclear coordinates and then transforming to mass-weighted
coordinates \cite{gaussian}. We have used the options corresponding to the low frequency vibrational
modes (opt=verytight, int=grid=ultrafine) in the GAUSSIAN09 program.

\section{Results and discussion}
In general terms, we have found qualitative similar results with the three functionals, particularly
all of them keep the same minimum vibrational state, except for the case of the configuration 11
that is shown in \Fig{1}.  The results shown in this paper are those with the functional B3LYP.

\subsection{Anthracene \& Nonacene}

Results for the vibrational modes of anthracene are shown in \Fig{1}. The different spectra are
ordered with increasing numbers of adsorbates starting from zero up to three for the configurations
our calculations show to be stable. In this figure and in all following vibrational spectra, we give
just the first 15 lower energy modes. Tight doublets are indicated by the letter D to make sure that
they can be identified. The actual movements for each configuration are shown in a set of short
movies available in the complementary materials. These movies are given for the lowest three modes
of all configurations shown in the figures with spectra, and they are identified as follows:
$n_1$-$n_2$-$n_3$, where $n_1$ identifies the number of the figure in this article, $n_2$ the column
in which the structure appears and $n_3$ the mode to which the clip belongs.

The first thing we notice in \Fig{1}, is that different numbers of adsorbates have different level
densities. We can then focus on situations with the same number of adsorbates and try to see whether
different configurations have markedly different spectra. This is indeed the case, and thus
vibrational spectra which are qualitatively correct, will identify the structure. The next step is
to see if the two, three and four adsorbate structures which have large deformation, i.e. the ones
with one or two benzene rings with two sided adsorption have any characteristic difference, as
compared to the other structures.  Here we must first draw attention to a case not previously
studied. In column 12 of \Fig{1} we see a skeleton with large deformation with three Li
adsorbates. The angle formed by the skeleton is actually even more pronounced then for the
previously studied situation of column 4. Furthermore we can see in the corresponding movie clip,
that the lowest vibrational mode is essentially only the Li at the apex performing a large
oscillation. A check of the charges, using the natural-bond-orbital analysis, shows that the
adsorbed Li are electropositive. The total transferred charge for the configurations of column 4 and
12 are roughly equal. In configuration 4 the Li at the apex and the Li inside the wedge transfer
almost the same charge.  In configuration 12 the Li at the apex transfers the same charge as the
corresponding Li in configuration 4, while the two Li inside the wedge transfer each almost the half
of it. In the movement of the different configurations we see, not unexpectedly, that some modes
involve essentially only movement of one or more adsorbates as seen best in the above example.

\begin{figure}[htb]
  \centering
  \includegraphics[scale=0.4]{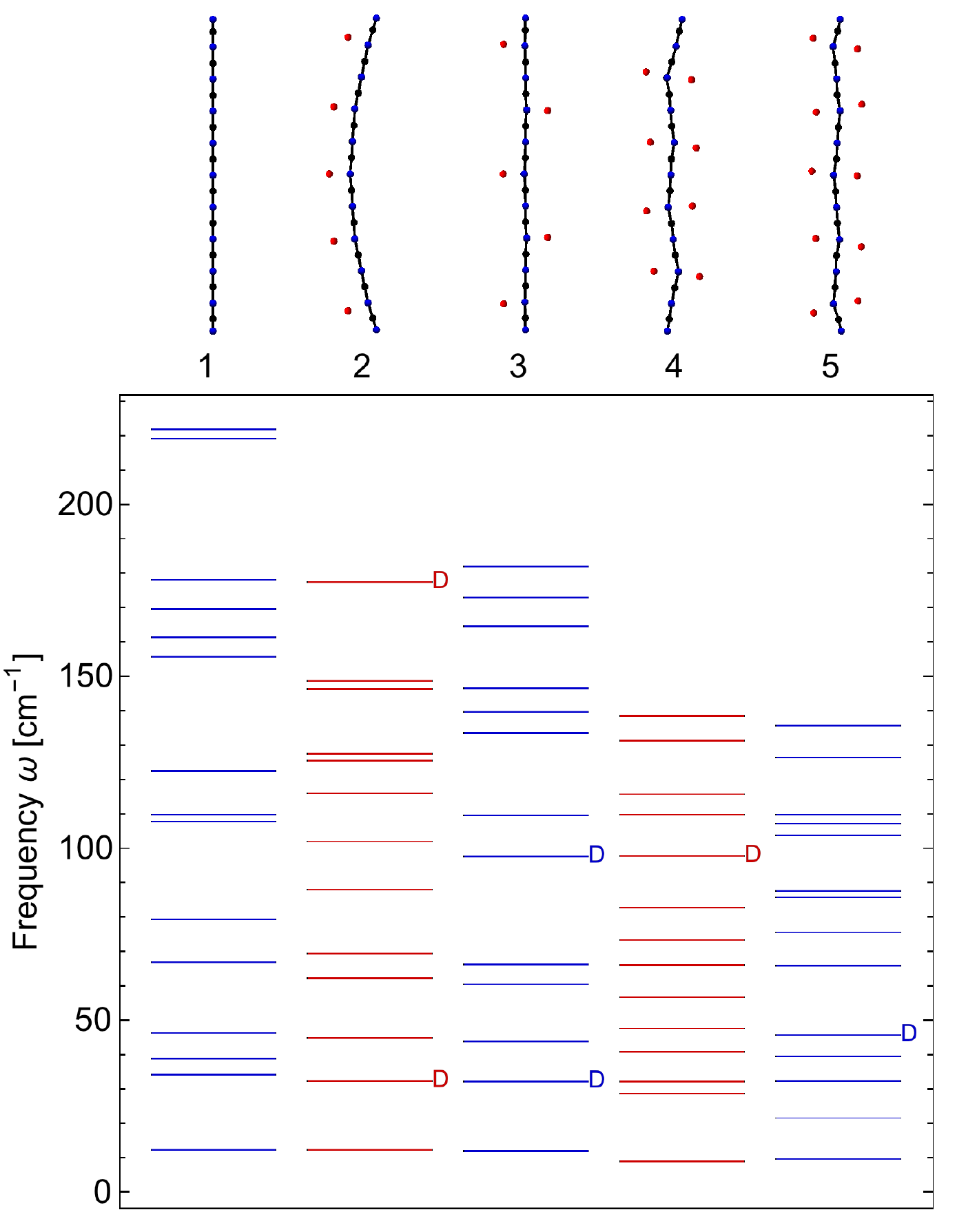}
  \caption{Vibrational spectra for nonacene and for some configurations of nonacene under Li
    adsorption.}
  \label{fig:2}
\end{figure}

After having shown the more obvious effects of adsorption on vibrational spectra one thing
emerged. There is a real chance for quasi-one-dimensional configurations which are also fairly flat
if the distribution of the adsorbates is more or less uniform along the string. In this case the
lowest modes mostly consist of an oscillation of the skeleton which is uniformly followed by the
adsorbates, which then do not much change their position relative to the benzene ring to which they
are adsorbed. To confirm our observations we look at selective situations of Li adsorption to
nonacene.  \Fig{2} already shows such four configurations for nonacene. In view of the wide field of
molecules we could extend this analysis to, a very hypothetical application may be considered to
guide us. The Innsbruck group of R. Blatt devised a quantum computer based on a quasi-1-D ion trap
as mentioned in the introduction. We shall look at the behavior of longer chains but restrict our
considerations to reasonably flat and symmetric configurations. Figure 2 already shows such
configurations for nonacene.  Again we see the lowest bending and torsional modes are of the type
where the strings with Li adsorption just follow the movements of the naked skeletons, making them
nice further examples we found in anthracene.


\subsection{poly-\textit{p}-phenylenes}

\begin{figure}[htb]
  \centering
    \raisebox{1.2cm}{$(a)$}\includegraphics[width=0.18\textwidth]{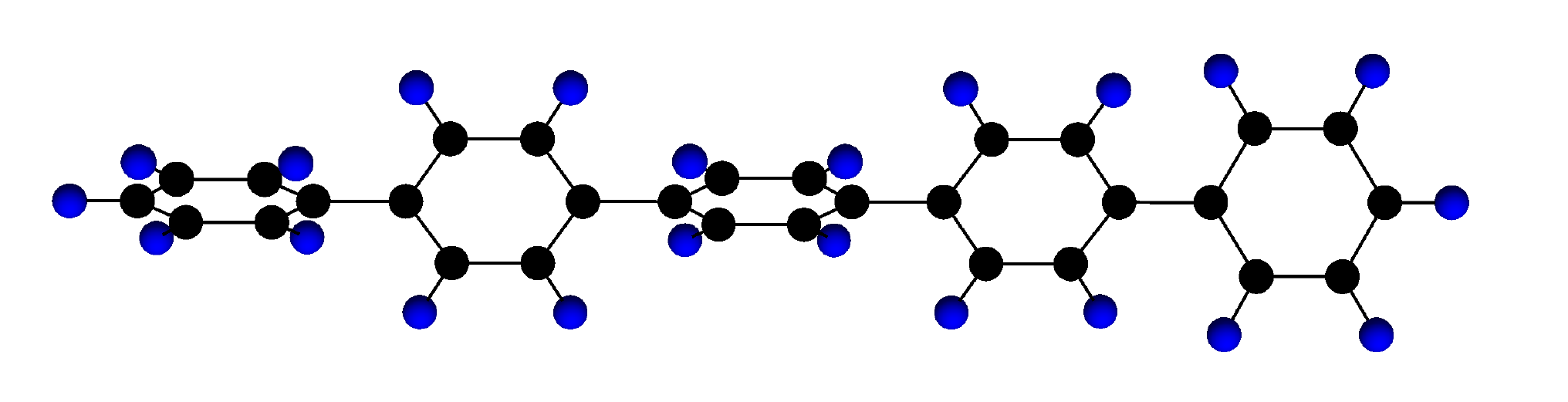}
    \raisebox{1.2cm}{$(b)$}\includegraphics[width=0.18\textwidth]{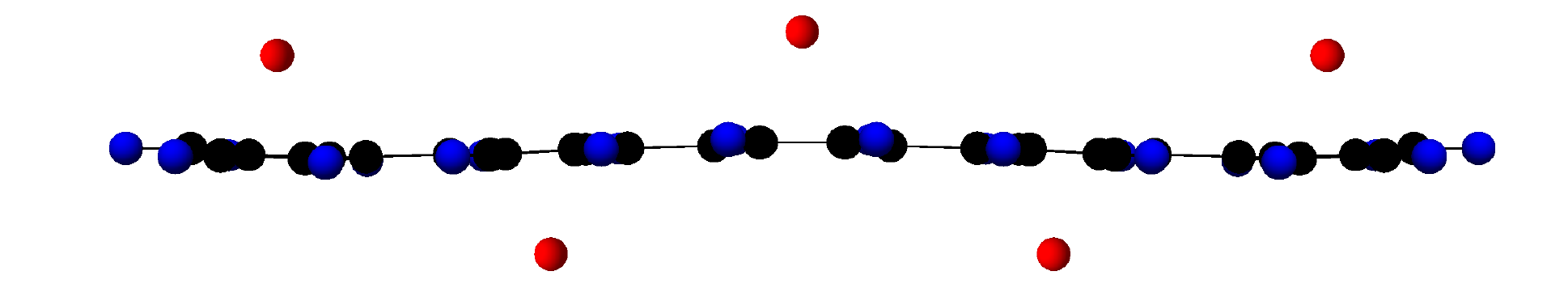}
    \raisebox{1.2cm}{$(c)$}\includegraphics[width=0.18\textwidth]{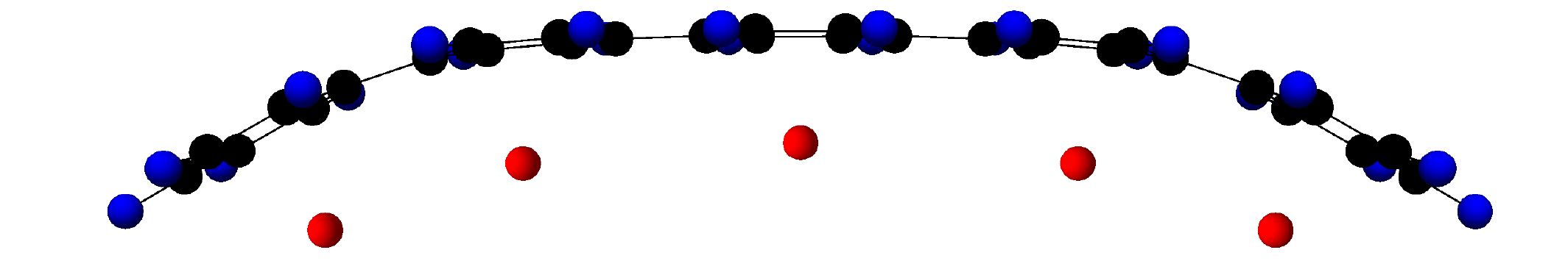}
    
    \caption{$(a)$ Pentaphenyl, $(b)$ Lateral view from adsorption of five lithium atoms alternating
      ring and side on pentaphenyl, $(c)$ Lateral view from adsorption of five lithium atoms on the
      same side on pentaphenyl}.
  \label{fig:3}
\end{figure}

\begin{figure*}[htb]
  \centering
  \includegraphics[scale=0.4]{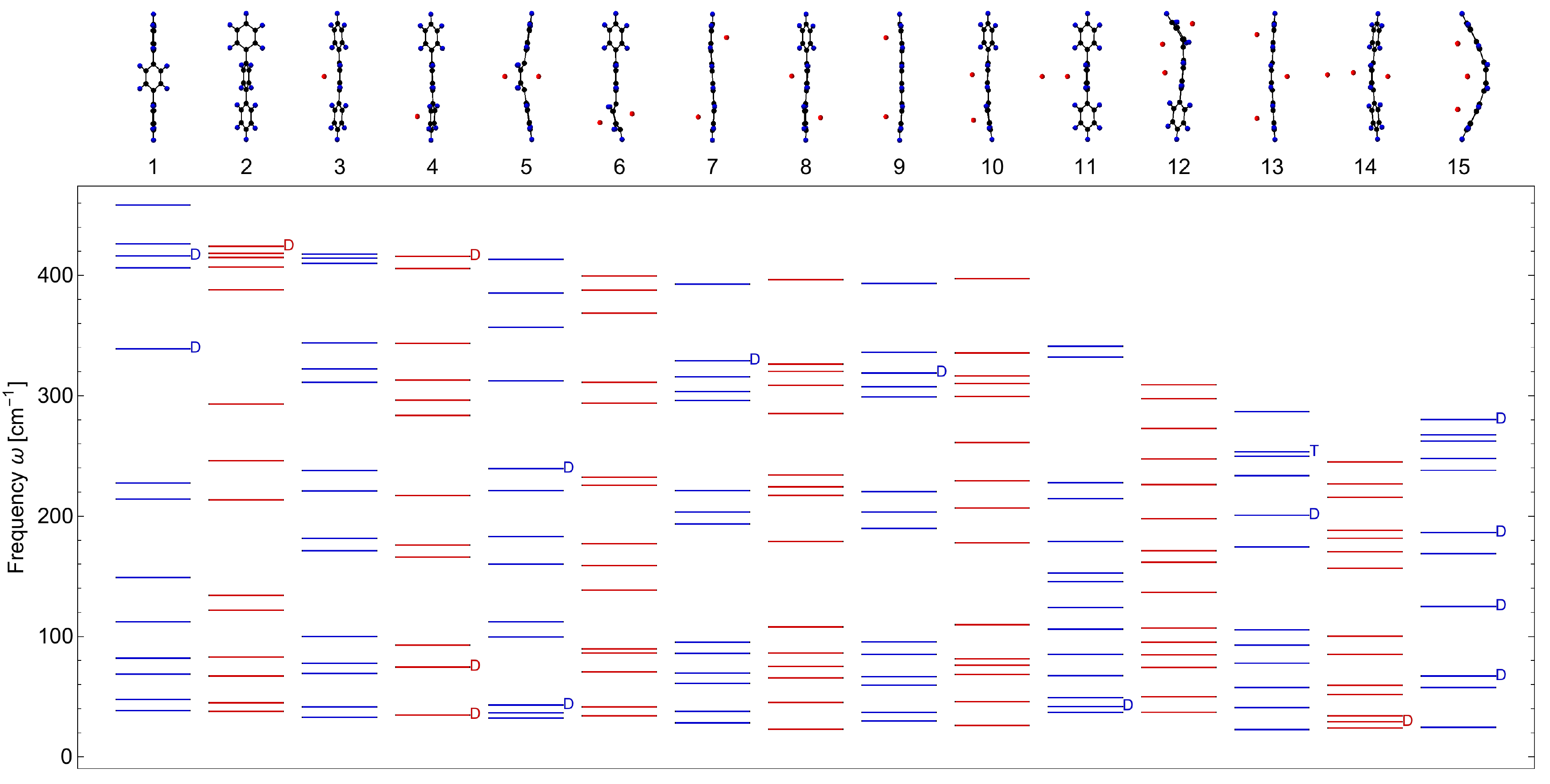}
  \caption{Vibrational spectra of triphenyl and triphenyl with Li adsorption. }
  \label{fig:4}
\end{figure*}

To carry this theme a little further let us try to be a little more realistic to deal with species
less reactive than the (higher) polyacenes, which tend to oxidize on exposure to ordinary
atmospheric conditions. Indeed this reactivity rapidly increases with polyacene chain length - so
that the outlook for long chains is much better if we look at phenylenes.  These are readily
available for large branchings if so desired, though we shall stick to linear chains or PPPs for the
present paper. An apparent disadvantage is the fact that the benzene rings often twist against one
another, at any chain length. Yet in \Fig{3} we see for pentaphenyl an interesting feature: both
single and double adsorption seem to eliminate this torsion to a large degree, at least if the
adsorbates cover the length of the PPP in an ordered manner. We can thus proceed to study vibrations
in several configurations obtained from DFT calculations.  In \Fig{4} we show selected
configurations of triphenyl under lithium adsorption. The different spectra are ordered with
increasing numbers of adsorbates starting from zero up to three for the configurations our
calculations show to be stable. The actual movements for each configuration are shown in a set of
short movies available in the complementary materials. Yet the question remains if Li in partial
ionization will provide an adequate two- level system and appropriate conditions can be
established. This cannot be the subject of this work, as it would go deeply into quantum optics.

The adsorbed Li atoms are electropositive, and so should give up electrons to leave
excess-electron-defected PPP. Then within a resonance-theoretic view it should be favorable for the
excess-$e$ resonating valence patterns to have the excess-$e$ delocalize between rings as would be
facilitated by enhancing inter-ring coupling by having neighbor rings aligned in a common plane. Of
course the effect of delocalization should fall off as the distance from the Li increases, so that
for sufficient number of ring to Li ratio, the twisting of rings should again set in (at the rings a
greater distance from Li). Moreover for Li-doped anthracenes delocalization of the excess-e should
tend to suppress the local distortion seen. The distinction of behavior between pentaphenyl and
anthracenes comes depending on how the resonance is enhanced for the anthracene by localizing the
defect in the center ring to leave 2 sextets in the terminal rings; whereas for pentaphenyl by
delocalizing the excess-e defect (through all different rings, it being seen that localization does
not in this case lead to any additional aromatic sextets)\cite{Klein2011,Clar1972}.
 
\begin{figure}[htb]
  \centering
  \includegraphics[scale=0.4]{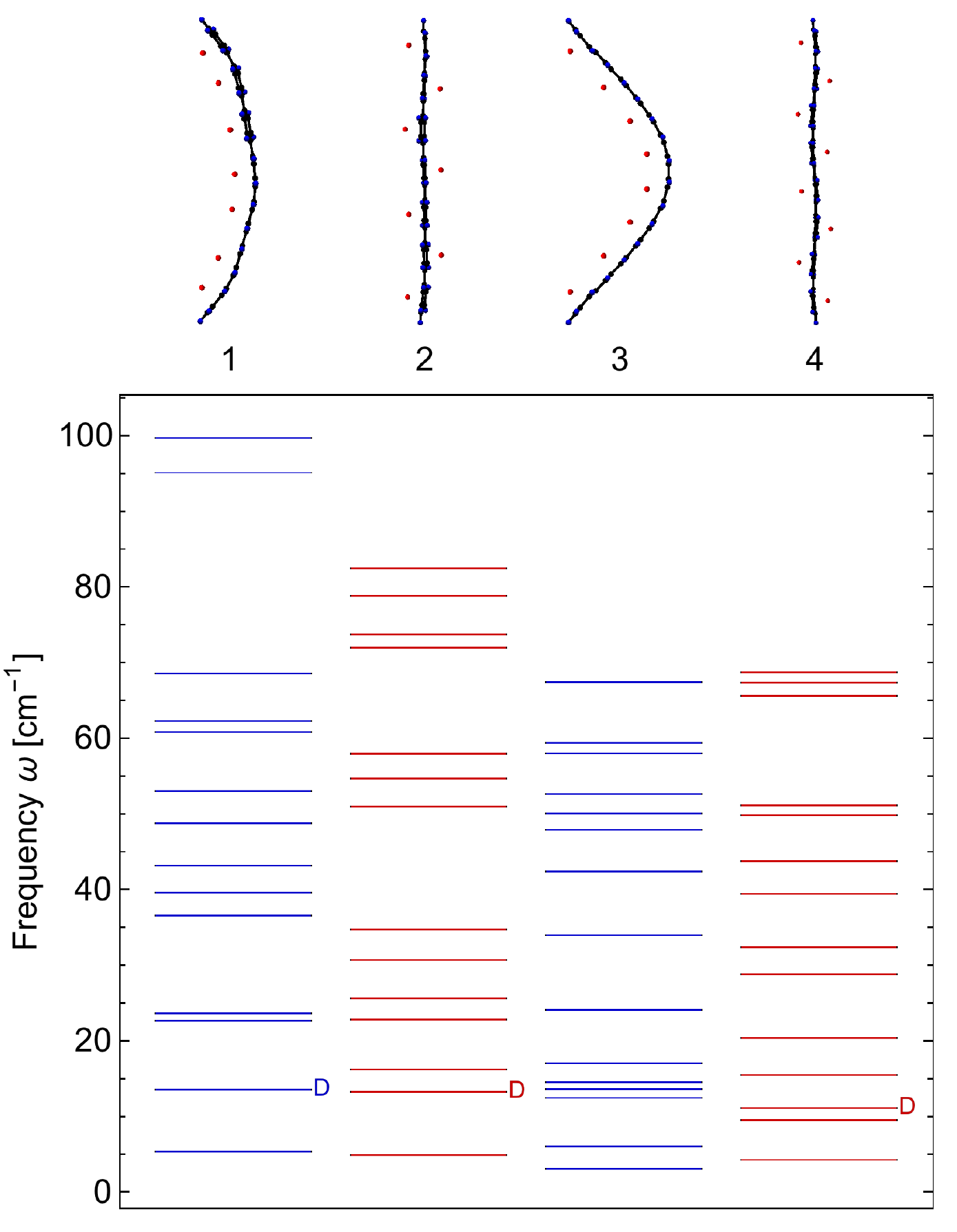}
  \caption{Vibrational modes of heptaphenyl  (column 1 and 2) and octaphenyl under Li adsorption  (column 3 and 4).}
  \label{fig:5}
\end{figure}

\begin{figure}[htb]
  \centering
    \raisebox{1.2cm}{$(a)$}\includegraphics[width=0.3\textwidth]{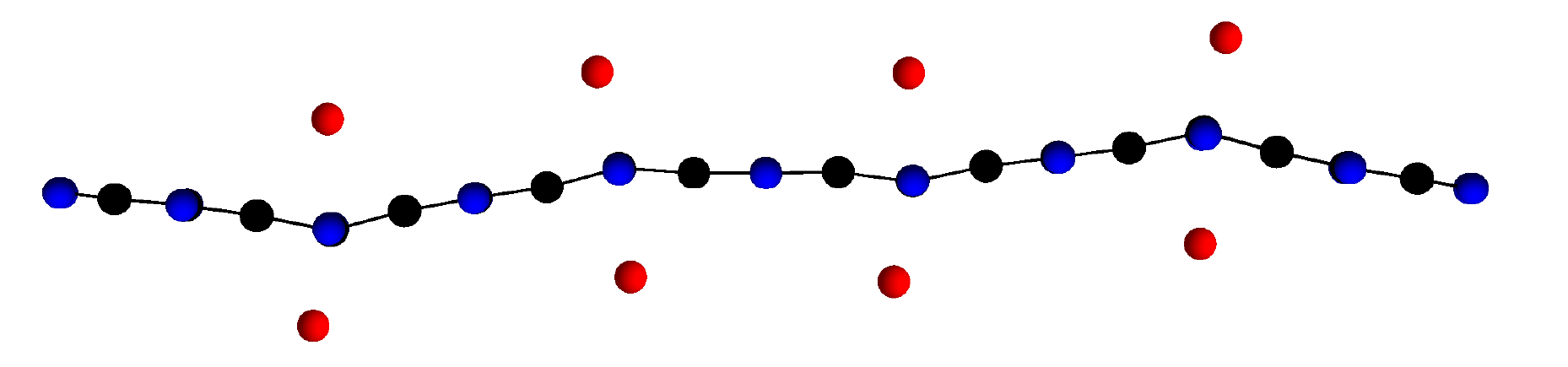}
    \raisebox{1.2cm}{$(b)$}\includegraphics[width=0.42\textwidth]{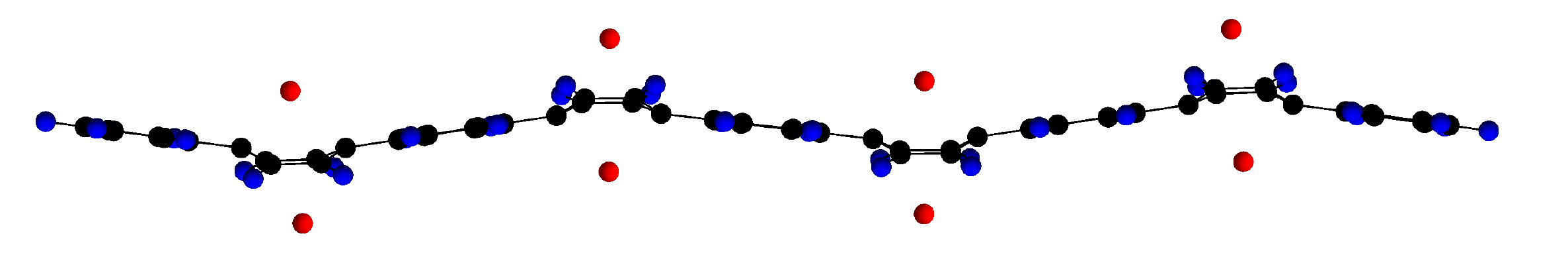}
    \caption{$(a)$ Lateral view from adsorption of four pairs of lithium atoms alternating ring on
      nonacene, $(c)$ Lateral view from adsorption of four pairs of lithium atoms alternating ring
      on nonaphenyl}
  \label{fig:6}
\end{figure}

To confirm the observation we show in \Fig{5} in the first two columns heptaphenyl with one and two
sided adsorption of 8 Li and equivalent configurations for octaphenyl are in the two columns. Indeed
the behavior is analogue to the one seen for tetraphenyl. We also attempted to calculate a
configuration with 16 Li adsorbed, eight on each side, but we found only irregular structures, as
had already been the case for tetraphenyl. We may note however that structures with double
adsorption to alternating rings are quite regular, though best applied to odd numbers of rings to
avoid a asymmetry between the ends of the chains. In this case we have found a spontaneous symmetry
breaking similar to what we have found for polyacenes resulting on a zigzag structure. It is shown
in \Fig{6} comparing nonacene and nonaphenyl where the inside angles are 160$^\circ$ and
164.6$^\circ$ respectively.  In infinite (or very long chains) a Peierls transition \cite{peierls}
is almost certain to occur.

\section{Conclusions}

In this paper we present a theoretical, largely numerical study of adsorption of lithium to
quasi-one dimensional conjugated carbon structures, namely polyacenes and poly-p-phenylenes
(PPPs). We analyze the vibrational spectra and qualitatively the movement of vibrational modes. We
have found that different adsorbate configurations lead to significantly different vibrational
spectra. The spontaneous symmetry breaking, which is observed for some adsorbate configurations,
does not show a characteristic signature in the vibrational spectrum.  Notably we find that the
torsion in the PPPs is greatly diminished and possibly eliminated by the alkaline adsorbates (shown
for lithium, but verified for other alkalines), if these are more or less regularly spaced. This
aspect is of importance. Polyacenes are theoretically very appealing, as the qualitative behavior of
the vibrational modes is similar. Yet as far as actual realizations are concerned PPPs have
practical advantages because they are less reactive. Thus we have similar structures for the two
types of molecules once they carry the corresponding adsorbates. We also speculate about
possibilities to use the deposited atoms as qubits and two modes of vibrations, as extended qubits
in order to mediate control gates.

\section*{Acknowledgments}
Financial support from CONACyT research grant 219993 and PAPIIT-DGAPA-UNAM research grants IG100616
and IG101113 is acknowledged.  We acknowledge extensive use of the MIZTLI super computing facility
of DGTIC-UNAM.

\appendix	

\section{Supplementary material}

The supplementary material contains movie clips of the lowest three vibrational modes of the studied
molecules. The movies are named as $n_1$-$n_2$-$n_3$, where $n_1$ is the number of the figure, $n_2$
is the number of the molecule indicated in the corresponding figure, and $n_3$ is the number of the
mode. For example, the movie clip 1-2-3 is the third mode of the second molecule in Figure 1.

The movie clips can be found also at \textbf{http://www.cicc.unam.mx/vibmod/v3}.

\section*{References}

\bibliographystyle{elsarticle-num}
  
\bibliography{vib-06-08-17.bib}

\end{document}